# From Prediction to Understanding: Will AI Foundation Models Transform Brain Science?

Thomas Serre and Ellie Pavlick
*Departments of Cognitive & Psychological Sciences and Computer Science*
*Carney Center for Computational Brain Science*
*Brown University*

## Abstract

Generative pretraining (the "GPT" in ChatGPT) enables language models to learn from vast amounts of internet text without human supervision. This approach has driven breakthroughs across AI by allowing deep neural networks to learn from massive, unstructured datasets. We use the term foundation models to refer to large pretrained systems that can be adapted to a wide range of tasks within and across domains, and these models are increasingly applied beyond language to the brain sciences. These models achieve strong predictive accuracy, raising hopes that they might illuminate computational principles. But predictive success alone does not guarantee scientific understanding.

Here, we outline how foundation models can be productively integrated into the brain sciences, highlighting both their promise and their limitations. The central challenge is to move from prediction to explanation: linking model computations to mechanisms underlying neural activity and cognition.

## Main text

Over the past year, two foundation models published back-to-back in Nature have drawn significant attention from brain scientists. The first is a neural foundation model trained on large-scale calcium imaging data from the mouse visual cortex[1]. The second, Centaur, is a behavioral foundation model trained to predict human decision-making across hundreds of psychology experiments[2]. Both achieve impressively predictive accuracy—generalizing across experimental tasks, subjects, and stimulus domains. These case studies sharpen a familiar question: what mechanisms, if any, do such predictors capture? We briefly review their training paradigms and then ask what predictive success can—and cannot—reveal about underlying mechanisms.

**The Pretrain-Finetune Recipe.** At the core of large language model (LLM) capabilities is self-supervised learning (SSL)—a family of approaches in which models learn by predicting missing or masked parts of their input without external labels. The most widely used form in LLMs is generative pretraining[3] (the "GPT" in ChatGPT), where models are trained autoregressively to predict the next token in large, unstructured text datasets, thereby acquiring broad linguistic and world knowledge. See Box 1 for additional forms of SSL and architectural details. A subsequent finetuning phase then adapts the model to specific domains and applications. In many cases, this involves supervised training on labeled datasets—for example, to classify clinical notes[4], analyze legal documents[5,6], or model political decision-making[7,8].

For open-ended text generation as in ChatGPT, models are first finetuned on human-written demonstrations—for example, prompts paired with ideal responses that illustrate helpful dialogue—and then further aligned using reinforcement learning. For example, in reinforcement learning from human feedback (RLHF), models optimize their responses to match human preferences rather than merely maximizing the likelihood of training text[9–12]. This two-step process effectively puts the "Chat" in ChatGPT, enabling models to generate responses that are more closely aligned with human intent. Variants of this RLHF pipeline are now standard across both general-purpose assistants and domain-specific systems, including mental health chatbots[13].

---

**Box 1 | Self-Supervised Learning**

Self-supervised learning (SSL) is a family of methods in which a model creates its own training signal. Rather than relying on external labels, the model withholds or corrupts parts of its input and learns to predict them from the surrounding context. This allows the system to learn directly from large, unlabeled datasets.

A widely used SSL approach is generative pretraining[3], introduced in the original GPT paper, where models are trained to predict the next element in an input sequence (e.g., the next word in a sentence). The basic units the model operates on are called tokens, which generalize across data types. In text, tokens may be words, sub-words, or characters; in images, they are patches; in audio or video, short frames or windows; and in neural or behavioral data, they can be time bins or events (e.g., spike counts). This shared tokenization principle allows the same SSL objectives to extend beyond language to images, audio/video, neural recordings, and behavior.

A foundation model refers to a network pretrained at scale (usually with SSL) on broad, heterogeneous data to learn versatile representations. These representations can then be adapted for downstream tasks using simple methods such as linear readouts, prompting, or finetuning.

Today's foundation models are almost all built on the transformer architecture[14], which has largely replaced earlier designs (MLPs, RNNs, CNNs). Transformers use a mechanism called self-attention, where each token's representation is updated based on weighted relations ("attention scores") with other tokens in the input sequence. In decoder-only models (e.g., GPT, Centaur[2]), attention is restricted to past tokens (causal masking), supporting next-token prediction. In encoder models (e.g., BERT[15] language model, calcium-imaging model[1]), attention is bidirectional, supporting masked-token prediction. Despite these differences, the common recipe is to pretrain a transformer on large datasets to learn general-purpose representations, and then adapt it to specific tasks.

For neuroscientists, a useful analogy is the contrast between hand-engineered features (e.g., filtering neural signals by frequency bands, applying PCA/ICA) and learned representations. Traditional workflows depend on manually defined features that are then fed into a decoder (often linear). SSL removes the manual feature-engineering step: the model learns features

---

> automatically from raw or lightly processed data at scale. Downstream analyses then often reduce to training a linear classifier on top of the learned representation.
>
> The remarkable progress of foundation models reflects this shift—from handcrafted features to learned, general-purpose representations—coupled with massive amounts of data and compute. This flexibility enables models to capture complex dependencies in neural, behavioral, and multimodal data, driving strong performance on predictive tasks[16].

SSL extends well beyond language. In vision, models learn by masking or removing image patches and predicting the missing content, enabling applications ranging from object recognition to medical imaging[17,18]. In speech, related methods reconstruct masked or corrupted audio segments, achieving state-of-the-art recognition with far less transcribed data[19]. In discrete domains such as programming, models trained on large code corpora predict missing or next tokens, supporting applications in code generation, translation, and debugging[20]. Across domains, the recipe is consistent: large-scale pretraining on raw data followed by application-specific finetuning. The same principles extend to multimodal models, which integrate text, images, audio, and even actions into shared representations. We summarize these developments in Box 2.

> **Box 2 | Foundation Model Scaling**
>
> Recent progress in AI has been driven by empirical scaling laws, which show that model performance improves predictably as data, compute, and parameters increase. These laws have guided the design of ever-larger foundation models pretrained on broad datasets and adaptable to diverse downstream tasks.
>
> The same pretraining–finetuning paradigm now enables multimodal learning across text, images, audio, and other data types. For example, Vision-language models[21,22] (VLMs) trained on large image–text datasets can support clinical clinical tasks, while vision–language–action[23] (VLA) models enable robots to follow natural-language commands.
>
> Frontier models—such as GPT-4o, Claude Sonnet 4, and Gemini 2.0—represent multimodal systems trained at unprecedented scales. Yet while scaling delivers impressive predictive gains, it does not guarantee that models uncover the causal mechanisms underlying the systems they aim to represent.

The same pretrain–finetune recipe is now being applied across the sciences, where foundation models are accelerating discovery in diverse domains. In biology, protein and genomics foundation models can predict 3D protein structures, design enzymes, and suggest functional mutations[24–26]—capabilities exemplified by AlphaFold[25], whose lead developers, Demis Hassabis and John Jumper, shared half of the 2024 Nobel Prize in Chemistry, with the other half awarded to David Baker for computational protein design. In climate science, AI-based weather models can forecast storms days in advance with accuracy that rivals the best physics-based systems[27,28]. In materials science, AI-based systems have proposed millions of candidate materials, many of which have already been synthesized and tested[29].

**Powerful Predictive Models for Neuroscience.** Just as AlphaFold transformed drug discovery by overcoming the long-standing bottleneck of protein structure prediction, neural foundation models promise to transform areas of neuroscience where accurate prediction is critical—for example, enabling adaptive deep brain stimulation for Parkinson's disease[30], identifying early behavioral or biological markers in depression or obsessive-compulsive disorders[31,32], or advancing neural prosthetics to restore walking after spinal cord injury[33].

A recent study introduced a foundation model of the mouse visual cortex trained on large-scale calcium imaging data[1]. The model not only predicts responses across new stimulus domains and individual animals with high accuracy, but also captures information about neuronal cell types, dendritic morphology, and connectivity. Similar progress is underway in human neuroimaging, where large fMRI foundation models predict brain states and clinical variables, while generalizing to new cohorts[34,35].

In cognitive science, Centaur is the first large-scale behavioral foundation model trained to predict human decision-making across a broad range of tasks and experimental settings[2]. The model encodes both natural-language task descriptions and prior participant choices as tokens, and predicts the participant's next choice from this sequence. Centaur outperforms classical cognitive models on held-out participants and generalizes to previously unseen task variations.

As neuroscience datasets expand, so too does the prospect of foundation model–based digital twins. For instance, high-density probes can now record from thousands of neurons across long timescales[36,37] and single-cell atlases aggregate tens of millions of cells with spatial context[38]. In principle, if a generative foundation model could produce neural or behavioral time series that are empirically indistinguishable (under rich tests) from those of an individual or cohort, it would function as a neurally realistic simulator. Such a simulator would form the core of a digital twin, enabling in silico experimentation, personalized medicine, and neurotechnology[39]. Yet indistinguishability is only a predictive criterion; it does not, by itself, provide a mechanistic account. Clarifying how digital twins might move beyond predictive mimicry toward mechanistic fidelity will determine whether they can advance neuroscience rather than remain another black box.

These examples sharpen the overarching question: do foundation models uncover causal mechanisms, or merely exploit statistical regularities? In AI, the issue is whether models recover any plausible generative process consistent with the data, or simply rely on pattern matching. In neuroscience and cognitive science, the challenge is sharper: do such models capture the specific data-generating process underlying neural activity and cognition, rather than just one of many fits? This distinction between prediction and explanation is long-standing, but the extraordinary predictive power of today's models makes it easy to mistake fit for understanding. In what follows, we review recent work that begins to address this central question in AI.

**Prediction is not Explanation.** There is growing optimism that, as foundation models improve, they may transition from capturing correlations to uncovering the generative processes underlying their data. If a model speaks like a human, perhaps it has internalized grammar; if it can play chess or Othello from game records alone[40–42], perhaps it has inferred the rules of the game. Such hopes are grounded in empirical scaling laws, which show that performance improves predictably with increasing model size, data, and compute.

Indeed, there is evidence that these hopes sometimes bear out. For example, while debates persist about the details, it is generally accepted that language models represent syntactic categories and relations, such as parts of speech and syntactic dependencies[43], and that these representations play a measurable causal role in the model behavior[44]. Recent work also suggests that transformer networks may capture learning dynamics that parallel those observed in humans. For instance, one study shows that the interplay between in-context and in-weight learning in neural networks mirrors dual-process theories in cognitive science, reproducing trade-offs between flexibility and retention and between blocked and interleaved curricula[45]. Together, these findings offer cautious optimism that foundation models might not only achieve predictive accuracy but also provide insights into the cognitive and neural mechanisms underlying language and learning.

At the same time, models can achieve strong performance not by capturing mechanisms but by exploiting spurious correlations—a phenomenon known as shortcut learning[46]. But even perfect predictive accuracy does not guarantee explanatory value: Ptolemy's epicycles, for example, offered accurate predictions of planetary motion while reflecting a false theory of celestial mechanics. Predictive abstraction can hold independently of mechanistic truth. It remains debated, for example, whether superhuman-level game-playing models trained purely on move sequences truly encode the game's rules[40–42,47,48], or whether models that solve analogies at a human level possess reasoning mechanisms like those of humans[49]. In both cases, studies show that performance degrades in situations not encountered during training.

A recent illustration comes from foundation models trained on synthetic orbital trajectories generated by Newtonian mechanics[50]. Although these models achieved high predictive accuracy, they failed to generalize to related physics tasks—suggesting that they had not internalized the underlying physical laws. This example highlights a key limitation: even when the true generative process is simple, well-defined, and embedded in the training data, a model may still fail to recover it. What evidence is there that current neural and behavioral foundation models capture genuine mechanisms of brain and cognition?

So far, the evidence remains limited. For example, the calcium-imaging foundation model[1] learns weight structures that organize information in ways partially consistent with anatomy—for instance, neuron types, dendritic morphology, and connectivity. Yet these correspondences fall short of revealing the circuit mechanisms by which cortex computes. Likewise, after finetuning, the Centaur behavioral model[2] shows increased alignment with human fMRI, linking its internal states to brain activity. However, critiques highlight that Centaur's predictions often diverge from human behavior in well-known psychology experiments[51]. In some cases, it achieves accurate performance even without access to task information, relying instead on statistical regularities in choice sequences—strategies fundamentally incompatible with human decision-making[52]. These limitations highlight the need for stronger criteria to assess when alignment between models and human data reflects genuine mechanism rather than surface-level fit.

These findings underscore that predictive alignment, on its own, is insufficient for mechanistic explanation. Without evidence that models have discovered genuine computational mechanisms of brain and cognition, we risk replacing one black box (the brain) with another (a deep neural network) of comparable complexity. Realizing the promise of foundation models for understanding cognition will require grounding them in established neuroscience and psychological theories, and identifying the specific mechanisms that underlie their predictive power. Only through such mechanistic understanding can foundation models generate testable hypotheses that genuinely advance our knowledge of human cognition.

**Toward Mechanistic Understanding.** This raises a practical question: how can foundation models be grounded in the theoretical frameworks of brain science? Conventional wisdom has long held that neural networks are "black boxes," antithetical to the kinds of explanations sought in neuroscience and psychology. Yet recent advances in interpretability are beginning to reveal computational structure at multiple levels.

The emerging field of mechanistic interpretability[53,54] seeks to map functional subcircuits that reproducibly implement specific computations. Analyses of attention weights and hidden-layer activations reveal specialized components—for example, circuits that extend sequences or suppress irrelevant inputs[55]. These elementary functions illustrate how complex behaviors emerge from combinations of simpler building blocks, which can be assembled into multi-step circuits for tasks like arithmetic[56] and factual recall[57].

Although these algorithmic "circuits" differ from biological ones, their compositional structure echoes theories of canonical microcircuits in neuroscience, where simple motifs are reused across the cortex[58]. Recent theoretical work[59] characterizes transformers in terms of primitive operations, offering the beginnings of a computational theory that predicts both the tasks LLMs can solve and the neural mechanisms that might support them.

At the representational level, research has uncovered individual model units with selectivity for particular concepts, echoing long-standing debates in neuroscience about "grandmother cells" and the merits of interpreting neural codes at the level of single neurons versus distributed populations. Early interpretability studies reported specialized units, such as "cat neurons" in large computer vision models[60], "gender neurons" in early language models[61], and more recently, a "Golden Gate Bridge feature"[62] (a distinctive learned pattern in the model's internal representation) in LLMs. This line of work parallels findings in human neuroscience, where single-unit recordings have revealed "concept cells" in the medial temporal lobe[63]—including neurons that respond selectively to landmarks such as the Sydney Opera House or to specific individuals such as Jennifer Aniston.

Many computations, however, are best viewed as distributed representations. Many LLM behaviors reduce to simple linear operations in high-dimensional activation spaces. Researchers can now "steer" language models through targeted interventions[64,65]. For example, given a neutral prompt such as the word "good" with no additional cues, the model can be directed to produce the antonym ("bad") by adding a vector pointing in the "antonym" direction, or to produce the Spanish translation ("bueno") by adding a vector pointing in the "Spanish" direction[66]. These steering directions generalize across contexts, suggesting that semantic relationships are encoded as consistent geometric patterns.

This points toward testable neuroscience hypotheses. If human conceptual knowledge follows similar organizational principles, then semantic relations should be recoverable through linear operations on neural activity patterns—a prediction testable with neuroimaging or electrophysiology. Importantly, this "linear algebra" hypothesis is only one among the earliest testable predictions to emerge from interpretability research. As interpretability research expands and our understanding of artificial neural representations deepens, many more concrete, experimentally testable hypotheses about brain function are likely to follow.

**Limitations and Opportunities.** Should we expect foundation models to converge on true brain mechanisms? In AI, scaling—training larger models on bigger datasets with greater compute—drives reliable predictive gains captured by empirical scaling laws. In biology, by contrast, increases in representational capacity emerge through evolutionary and developmental constraints. This asymmetry helps explain why empirical scaling laws may yield predictive gains without converging on the path-dependent mechanisms of biology.

Approaches that embed evolutionary- and development-like constraints offer one path to bridge the gap between prediction and mechanistic explanation. For example, researchers can begin with biologically grounded architectures and test how optimization yields explanatory insights[67], or apply deep learning to identify developmental principles shaping visual representations[68]. We develop this argument in more detail elsewhere, highlighting how the absence of evolutionary constraints may limit the explanatory value of current AI models[69]. Such strategies may point the way toward foundation models that are not only predictive but also mechanistically explanatory.

But even if the mechanisms discovered in foundation models differ from those in the brain, interpretability can yield theoretical insights. Identifying computational mechanisms within these models and generating testable hypotheses fuels the crucial feedback loop between theory and experiment, driving scientific progress. Whether or not the mechanisms are "correct," this process advances our understanding of both artificial and biological intelligence.

**Conclusion.** Foundation models alone will not transform neuroscience. Their scientific value depends on moving beyond prediction to interpretation—understanding their computations, grounding them in brain sciences theory, and designing experiments to test those links. Digital twins may become powerful tools, but their value hinges on converting predictive structures into mechanisms grounded in brain science. As Ptolemy's epicycles remind us, even perfect prediction is no substitute for mechanistic explanation. The future of neuroscience with foundation models depends on whether we can transform data-fitting machines into theory-bearing scientific instruments—tools that reveal not only what intelligence can accomplish, but also how it works.

# Acknowledgements

We would like to thank Wael Assad, David Badre, John Davenport, Alexander Fleischman, Mikey Lepori, Drew Linsley, Maria Grazia Ruocco, and Gretchen Schrafft for feedback on the manuscript. T.S. was supported by the Office of Naval Research (N00014-24-1-2026 and REPRISM MURI N00014-24-1-2603), the National Science Foundation (IIS-2402875 and EAR-1925481), and the ANR-3IA Artificial and Natural Intelligence Toulouse Institute (ANR-19-PI3A-0004). E.P. was supported by a Young Faculty Award from the Defense Advanced Research Projects Agency (Grant #D24AP00261).

# Declaration of Interest

Infant Egocentric Videos.